\documentclass{amsart}

\usepackage{amsmath}
\numberwithin{equation}{section}
\DeclareMathOperator{\RE}{Re}
\newcommand{\z}{\mathbb{Z}^+}
\newcommand{\ta}{IIA }
\newcommand{\D}{\bar{D}}
\newcommand{\pd}{\partial }

\begin{document}

\title[Quantum Kaluza--Klein Compactifiation]
 {Quantum Kaluza--Klein Compactifiation}

\author{ Corneliu Sochichiu }

\address{Bogoliubov Laboratory of Theoretical Physics\\
Joint Institute for Nuclear Research\\ 141980 Dubna, Moscow Region
\\ Russia}

\email{sochichi@thsun1.jinr.ru}

\thanks{This work was completed with the support of RBRF grant
96-01-0551, Scientific School Support grant 96-15-96208, and INTAS grant
96-370.}

%\date{}

\dedicatory{}

%%% ----------------------------------------------------------------------

\begin{abstract}
Kaluza--Klein compactification in quantum field theory is analysed
from the perturbation theory viewpoint. Renormalisation group
analysis for compactification size dependence of the coupling constant is
proposed.
\end{abstract}

%%% ----------------------------------------------------------------------
\maketitle
%%% ----------------------------------------------------------------------

\section*{Introduction}

Development of the string theory \cite{gsw} increased interest to
field theoretical and quantum mechanical models in high dimensions
($D\geq 5$). The consistency of string theory need it to be
formulated in either $D=10$ for supersymmetric strings or $D=26$
for bosonic ones, while ``everyday'' physics is four-dimensional.
From the other hand most interesting field theoretical models can
be consistently formulated at dimensions not exceeding four.
Compactification of extra dimensions existent in string theory to
have four-dimensional physics at the low energies conciliates
these facts.

It would be interesting to explain this mechanism dynamically, at
least in some approximation at low energies. Much optimism is
inspired by the progress in understanding non-perturbative strings
and and in special ADS/CFT correspondence \cite{mald,pol,wit}.

In the present work, however, we address a different question:
what is the effect of the compactification on the level of quantum
field theory?

It is known that certain quantum field theories/gravities can
serve as low energy effective actions for string theories. Since
the string description must not enter in contradiction with low
energy field theoretical one, one can limit oneself to study of
the last.

As an important example can serve ten-dimensional \ta string model
whose low energy field theory is \ta supergravity. As it is
believed string theory for large couplings ($g_{string}$) results
in an eleven-dimensional model (M-theory). Its low energy
effective field theory model is $D=11$ supergravity. Therefore,
stringy corrections for large $g_{string}$ in \ta supergravity
must led to the eleven-dimensional theory (see \cite{kir} for a
recent review). It is, however, easer to see instead the
transition from higher dimensional model to lower one.

In classical physics, compactification of a $(D+p)$-dimensional
model to $D$ dimensions is given by ``confining'' some $p$ spacial
dimensions of original space-time to form a compact manifold.

From the $D$-dimensional point of view the spectrum of the
compactified model consists of a light field which corresponds to
constant or zero mode on the compact directions and an infinite
number of massive fields corresponding to non-constant in the
compact direction modes or Kaluza--Klein (KK) modes. Masses of
KK-modes are proportional to inverse compactification size
$M=R^{-1}$ (i.e. the typical size of the compact dimensions). In
the limit of strong compactification KK-modes acquire large masses
and do not propagate, and decouple at the classical level.

In quantum description, however, ``virtual'' KK-particles can
contribute even for energies less than their masses, becoming
significant as they are approached.

In the limit of zero compactification size one expects to have all
the KK-modes decoupled also in quantum theory since their masses
become infinite. In fact, the fields not only do not simply
decouple in this limit but they may also produce additional
divergences. Also, extra divergences appear even for finite
compactification sizes due to infinite number of fields. These
divergences are natural reflection of the fact that usually
similar models in higher dimensions are more divergent.

As one can see  these divergences can be eliminated in the
framework of the standard renormalisation procedure and one is
left with renormalised physical parameters which depend on the
size of the compact space.

In actual paper we are going to consider a simple (toy) model to
illustrate above ideas. We will consider $D+1$-dimensional
$\phi^3$-model compactified (on a circle) to a $D$-dimensions.

The structure of the paper is as follows. In the next section we
consider compactified $D+1$-dimensional $\phi^3$ model, and in the
second one its $D$-dimensional one-loop effective action. We also
analyse the renormalisation group dependence of the effective
$D$-dimensional coupling on the compactification mass $M$. In the
Appendix we give some properties of $\zeta$ and $\Gamma$-functions
used in the body of the paper and describe the computation of the
effective action.
%%% ----------------------------------------------------------------------
\section{Compactified $\phi^3$ model}
Let us consider $D+1$ dimensional $\phi^3$ model described by the
following classical action
\begin{equation}\label{d+1}
  S=\int d^{D+1}\bar{x} \left(\frac12 (\partial_{\bar{\mu}} \bar{\phi})^2-
  \frac12 m^2\bar{\phi} ^2-\frac{\bar{\lambda}}{3!}\bar{\phi}^3\right),
\end{equation}
where bar always refer to $D+1$-dimensional quantities. We
compactify this model along $D$-th spatial direction by requiring
equivalence of $x^{D}\equiv\theta \simeq\theta+2 \pi R$, where
$R\equiv M^{-1}$ is the size of compactification.

Consider the (infinite) set of $D$-dimensional fields $\phi_n (x)$
which is Fourier transform with respect to the $D$-th coordinate
$\theta$

\begin{equation}\label{d+1->d}
  \bar{\phi}(\bar{x})=(2\pi R)^{-1/2}\sum_{n=-\infty}^{+\infty} \phi_n(x)
  e^{iMn\theta},
\end{equation}
where $M=R^{-1}$ is the energy scale of compactification, and
fields $\phi_n$ are given by inverse Fourier transform
\begin{equation}\label{d->d+1}
  \phi_n(x)=(2\pi R)^{-1/2}\int_0^{2\pi R}d\theta \, \bar{\phi}(\bar{x})e^{-iMn\theta}.
\end{equation}

In terms of fields $\phi_n$ action (\ref{d+1}) look as follows
\begin{equation}\label{d_action}
  S= \int_{M_D} d^Dx \left( \sum_{n\geq 0}
  \left(\frac12\partial\phi_n\partial\phi_n^*-\frac12
  (m^2+M^2n^2)\phi_n\phi_n^*\right)-\frac{\lambda}{3!}
  \sum_{n,n'}
  \phi_n\phi_{n'}\phi_{n+n'}^{*}\right).
\end{equation}

The $D$-dimensional coupling $\lambda$ in eq.(\ref{d_action}) is
related to $D+1$-dimensional one $\bar{\lambda}$ by
compactification size dependent rescaling:
\begin{equation}\label{lambdas}
  \lambda=\frac{\bar{\lambda}}{\sqrt{2\pi
  R}}=\sqrt{\frac{M}{2\pi}}
  \bar{\lambda}
\end{equation}

The rescaling take place due to dependence of coupling
dimensionality on the space-time dimension. Indeed, dimensions of
the scalar field $\phi$ and cubic coupling $\lambda$ in $D$
space-time dimensions are respectively (in mass unities):
\begin{eqnarray}\label{dphi}
[\phi]=\frac{D}{2} -1,\\ \label{dlambda} [\lambda]=3-\frac{D}{2}.
\end{eqnarray}
Thus, the coupling must acquire a $\sim M^{1/2}$ factor while
descending one dimension.

Now, the the action for zero-mode field $\phi(x)\equiv \phi_0 (x)$
only is just the naive $D$-dimensional scalar field action, i.e.
one that one would have in $D$ dimensions. Beyond this standard
$D$-dimensional part there are also terms for higher KK-modes
$\phi_n$, ($n \neq 0$) and ones responsible for their interaction
with $\phi$. Making such separation in $D$-dimensional fields and
KK-modes explicit one can rewrite the action (\ref{d_action}) in
the following form
\begin{equation}\label{explicit}
  S_{D+1}=S_{D}(\phi )+S'(\phi , \phi_n,\phi_n^{*}),\quad n>0
\end{equation}
where actions $S_{D}(\phi )$ and $S'(\phi , \phi_n,\phi_n^{*})$
are given by the following $D$-dimensional Lagrangians
\begin{equation}\label{sdphi}
  L_{D}(\phi )=\frac12(\partial\phi)^2-\frac12 m^2
  \phi^2-\frac{\lambda}{3!}\phi^3
\end{equation}
and
\begin{equation}\label{int}
  L'=\sum_{n>0}\left(\frac12|\partial\phi|^2-\frac{M^2n^2}{2}|\phi_n
  |^2-2\lambda \phi_n \sum_{n>0} |\phi_n|^2 \right) +
  \frac{\lambda}{3!}
  \sum_{n,n'}{}' \phi_n \phi_{n'} \phi_{n+n'}^*,
\end{equation}
where the primed sum is taken over all values of $n$ and $n'$
which satisfy $n,n'\neq 0 $ and $n\neq n'$.

{}From the explicit form of eqs (\ref{explicit}-\ref{int}) one can
see that fields $\phi_n$, $n\neq 0$ are charged with respect to
$U(1)$ group acting as $\phi_n\rightarrow e^{-i\alpha n} \phi_n$.
This transformation corresponds to shifts (rotations) in $D$-th
(compact) direction, and leaves the action (\ref{explicit})
invariant. Gauging this symmetry give rise to $U(1)$ KK gauge
field ${\mathcal A}_\mu $.

In the compactification limit: $R\rightarrow 0$ ($M\rightarrow
\infty $) the KK modes aquire infinite masses and as we mentioned
are expected to decouple. In this limit KK modes do not propagate
anymore, but due to their interaction with remaining
$D$-dimensional fields they can produce a non-vanishing
contribution. In fact this contribution is divergent, divergencies
being accomplished also by the infinite number of KK-fields.

The extra divergences of compactified model are easily explained
by the fact that index divergence in higher dimensions is worse
than in lower.

To evaluate the effect of compact KK-modes on $D$-dimensional
theory let us compute the effective action for zero mode $\phi $,
in one-loop approximation.

\section{The Effective Action}
The effective action for the field $\phi $ is given by the
following equation:
\begin{equation}\label{eff}
  e^{iS_{eff} (\phi )}=e^{iS_D}\int \prod_{n>0}d\phi_n d\phi_n^* e^{i\int
  d^D x\,L'(\phi,\phi_n,\phi_n^*)}.
\end{equation}

In what follows we will consider compactification size to be
small. The presence of the compactification size in the model
introduces a new scale parameter. In fact one can identify this
scale with the cut off one, but we will not do this at the moment.

To compute effective action $S_{eff}$ at least in the framework of
the standard perturbation theory one needs first to regularise
path integral (\ref{eff}). During this calculation we use
dimensional regularisation scheme and perform Wick rotation:
$x_0\rightarrow i x_0$, to deal with Euclidean path integral.

KK-mode propagators  look as standard Euclidean scalar propagators
(in momentum representation):
\begin{equation}\label{prop}
  D_n(p)=\frac1{p^2+M^2n^2}.
\end{equation}
There are also two interaction terms. The first one is
$\phi$-KK-mode interaction:
\begin{equation}\label{phikk}
  2\lambda \phi \sum_{n>0}|\phi_n|^2,
\end{equation}
and the second one is KK-mode self-interaction term:
\begin{equation}\label{kksi}
  \lambda\sum_{n,n'\neq 0} \phi_n\phi_{n'}\phi_{n+n'}^*.
\end{equation}

Since we are considering one-loop approximation only the
$\phi$-KK-interaction (\ref{phikk}), is relevant.

Typical one-loop diagram with $N$ ``legs'' $\phi (q_i)$
($i=1,\dots , N$), produces the following regularised
contributions\footnote{As usual in dimensional regularisation
computations we assume the coupling to be of the form
$\lambda=\lambda_0 \kappa^{3-D/2}$, where $\kappa$ and $\lambda_0$
are respectively a mass dimensional ``unity'' and dimensionless
coupling.} (see Appendix A):
\begin{equation}\label{fc}
  G_N=\frac{\lambda^N}{(2\sqrt{\pi})^{\D
  }(N-1)!}M^{\D-2N}\left[f_N^{(0)} (\D)+M^{-2} f_N^{(2)}(\D
  ;q_1,\dots q_N)  +O((q/M)^4)\right],
\end{equation}
where $\D \rightarrow D$ is the (complex) dimension which
regularises the theory.

Functions $f_N^{(i)}$ in eq. (\ref{fc}) have the following
structure,
\begin{equation}\label{f}
  f_N^{(i)}=\zeta\left(2N+i-\D \right)\Gamma \left(N+\frac{i-D}{2}
  \right)P_{(i)}(q),
\end{equation}
where $P_{(i)}(q)$ is a (homogeneous) polynomial of the $i$-th
degree in external momenta $q_l$, $l=1,\dots,N$.

As one can expect, the right side of eq. (\ref{fc}) is divergent
in the limit $\D \rightarrow D$. Let us analyse this divergence
and find the counter-terms necessary for its cancellation.

The UV divergencies in eq. (\ref{fc}) manifest themselves as a
potential singularity of the factor
\begin{equation}\label{singg}
  \zeta(2N+i-\D)\Gamma
\left(N+\frac{i-\D}{2}\right)
\end{equation}
as $\D$ goes to $D$.

From the properties of $\zeta$- and $\Gamma$-functions (Appendix
B) one can deduce that the singularity in eq. (\ref{singg}) occurs
when quantity $N+\frac{i-D}{2}$ is either $0$ (then one has
singularity in $\Gamma$-function times regular $\zeta$) or
$\frac12$ (when, oppositely, one has regular $\Gamma$ times
singular $\zeta$). As one can see, the latter of these two cases
can be met for odd dimension $D$ while the former happens when
dimension is even.

Concerning the compactification mass value there may be two
essentially different situations. The first one is when the
compactification mass is below the cut-off scale, then a
physically meaningful value can be fixed for it. In the second
situation the compactification mass is beyond  the cut-off it is
physically infinite and one meets additional renormalisable
divergences due to $M\to \infty$. Their elimination by standard
renormalisation of fields, masses, couplings, and mean vacuum
field value brings us to usual scalar field model and one cannot
speak on the compactification size dependence since it is
physically infinite. We are interested mainly in the first
situation.

In the case when the compactification mass $M$ is kept fixed below
the cut-off scale the structure of the UV divergencies look as
follows,
\begin{equation}\label{diverg}
    \Delta
    Z_N^{(i)}=\frac{\lambda^N}{(2\sqrt{\pi})^{D}(N-1)!}P_{(i)}(q)
    \cdot \left(\frac{1}{2\varepsilon}\right)+\text{finite terms},
\end{equation}
$N+\frac{i-D}{2}=0$, for even dimension $D$, and, respectively,
\begin{equation}\label{divergodd}
     \Delta
    Z_N^{(i)}=\frac{\lambda^NM^{-1}}{(2\sqrt{\pi})^{\D}(N-1)!}P_{(i)}(q)
     \cdot \left(\frac{\sqrt{\pi}}{\varepsilon}\right)+\text{finite terms},
\end{equation}
$N+\frac{i-D}{2}=\frac12$, for odd one, also
$\varepsilon\equiv\D-D$.

In what follows let us consider the coupling constant
renormalisation due to presence of the compact extra dimension. In
our notations the coupling $\lambda$ renormalises as follows,
\begin{equation}\label{ren}
  \lambda_R= \lambda Z_3^{(0)}(Z_2^{(2)})^{-3/2}.
\end{equation}
As one can immediately see the coupling acquires an infinite
renormalisation only in dimensions $D=5$ and $D=6$. In other cases
both $Z_3^{(0)}$ and $Z_2^{(2)}$ are finite.

The case of five dimensions is particular due to compactification
mass dependence of divergent terms. This leads to compactification
mass dependence of the renormalisation procedure. As a result one
cannot define in five dimensions ``physical'' coupling for all
values of $M$ simultaneously, but must re-renormalise it for each
particular value of $M$.

In other dimensions ($D\neq 5$) the renormalisation is either
finite or it is compactification mass independent, as it is in
$D=6$.

Consider now, dimensions different from $D=5, 6$. For dimensions
four and less, eq. (\ref{ren}) is regular and one just has,
\begin{equation}\label{less}
  \lambda_R=
  \lambda\left(1-\frac{\lambda^2}{4(2\sqrt{\pi})^D}M^{D-6}
  \zeta(6-D)\Gamma(3-D/2)\right).
\end{equation}
The same is true (and regular) also for odd dimensions greater
than six. For even dimensions greater than six, however, eq.
(\ref{less}) has singularity in $\Gamma$-function which is
cancelled by zeroes of $\zeta$-function. Computation the limit
yields,
\begin{equation}\label{twon}
  \lambda_R =\lambda\left(1-\frac{(-1)^n\lambda^2}{4(2\sqrt{\pi})^D
  n!}M^{D-6}\zeta' (6-D) \right),\qquad D\equiv 2n>6
\end{equation}

Consider now in more details the most interesting case $D=6$,
where the model is yet renormalisable but not superrenormalisable.
Renormalisation group equation gives\footnote{Here and on we drop
subscript $R$ at $\lambda$ since the only $\lambda$ we deal with
is $\lambda_R$.},
\begin{equation}\label{rengr}
  M\frac{\pd \lambda}{\pd M}=\beta (\lambda),
\end{equation}
where $\beta (\lambda) $ is Calan--Simanzik $\beta$-function. It
can be computed from the equation,
\begin{equation}\label{betaf}
  \beta(\lambda)=\kappa \frac{\pd \lambda}{\pd \kappa}.
\end{equation}

Using eqs. (\ref{ren}) and (\ref{fc}) one has for
$\beta$-function,
\begin{equation}\label{sixbeta}
   \beta (\lambda)=-\frac{\lambda^3}{(2\sqrt{\pi})^6}.
\end{equation}

Thus solution to (\ref{rengr}) yields,
\begin{equation}\label{sol}
   \lambda^2 (M)=\frac{\lambda^2
   (M_0)}{1-\frac{2\lambda^2(M_0)}{(2\sqrt{\pi})^6} \log (M/M_0)},
\end{equation}
where $M_0$ is the value of the compactification mass parameter
where $\lambda (M_0)$ was computed.

Eq. (\ref{sol}) gives the dependence of the effective coupling
$\lambda$ on compactification mass parameter $M$. This equation is
valid in the approximation $p\ll M$. The behaviour of the coupling
is of such nature that it exhibits a singularity at $M\to \infty$
(or compactification size $R\to 0$).

This singularity could be expected since there are additional
logarithmic divergences of contributing Feynman diagrams as $M$
goes to infinity.

As a conclusion one cannot compactify seven-dimensional scalar
field model to six-dimensional one in a way smooth in the
compactification size. Renormalisation group behaviour shows that
at small values of $R=M^{-1}$ the coupling constant rises
disabling the perturbative analysis.

This is in contrast with the so called large mass decoupling
problem \cite{appcar,wi,wei,os} (see also \cite{col})\footnote{I
am grateful to Dr.~R.~Ruskov, who pointed my attention to this
problem.}.

 From the other hand, there maybe such a situation for special
choice of model and compact manifolds when the coupling's
behaviour is asymptotically free at large $M$, in this case one
can safely reach the compactification limit.

\subsection*{Discussions} We found the compactification mass
(inverse of compactification size) dependence of the
$D$-dimensional coupling of the model resulting in
compactification of the $D+1$ dimensional scalar field model. We
considered the simplest possible case: scalar field and $U(1)$
compactification. However, this work can be extended to more
complicated cases both as field content and as the type of
compactification, including various compactifications of $D=11$
and $D=10$ supergravity theories~\cite{workinpr}.

More complicated models may contain fields which realise
non-singlet representations of the $D+p$-dimensional Lorentz
group, which must be reduced under compactification to
representations of $D$ dimensional Lorentz group. This reduction
was considered in the famous Slansky's review~\cite{slan}.

It is important that compactified model may be field theoretically
consistent in spite the non-renormalisability  of the original
non-compactified one. Therefore one may expect a phase transition
with compactification size as order parameter where the model pass
to the $D+p$-dimensional phase. This is exactly the point where
the $D$-dimensional perturbation theory fails. Thus one may think
about $D+p$-model as a non-perturbative extension of the
$D$-dimensional one. As we have shown on the toy model considered
in the paper  the renormalisation group approach can be applied to
study the compactification size dependence of the model.

\subsubsection*{Acknowledgements}
I wish to thank the Institute for Atomic Physics (Bucharest),
where the main part of the present work was done, for hospitality.
I am specially obliged to Profs. G.~Adam, S.~Adam and Dr.
\c{S}~Mi\c{s}icu as well as to Prof. M.~Vi\c{s}inescu for
attention paid to my work and for useful discussions.
%-------------------------------------------------------------------------
%-------------------------------------------------------------------------
\appendix
\section{Computation of the Effective Action}

One-loop Feynman diagram with $N$ external legs $\phi (q_i)$
produces the following contribution to effective action:
\begin{eqnarray}\label{feyn}
&&G_N(q_1,\dots ,q_N)=
  \frac{\lambda^N}{N!}\sum_{n>0}\int
  \frac{d^{\D} p}{(2\pi )^{\D} } \times \\
  \nonumber
  &&\times [((p^2+M^2n^2)((p+Q_1)^2+M^2n^2)\dots
  ((p+Q_{N-1})^2+M^2n^2)]^{-1} \\
  \nonumber
  &&\qquad +\text{permutations}(q_1,\dots,q_N),
\end{eqnarray}
where
\begin{equation}\label{Q}
  Q_i\equiv \sum_{k=1}^{i}q_k
\end{equation}

In the framework of dimensional regularisation scheme, $\D$ is the
complex dimension. The limit of ``cut-off'' removing is $\D =D$.
As usual, $\lambda=\lambda_0 \mu^{N(\D/2-3)}$ where $\lambda_0$ is
dimensionless coupling constant and $\mu$ is the mass scale.

Since we are considering masses $M$ to be large we can expand
integral (\ref{feyn}) in powers of $q/M$. Due to Lorentz
invariance only even power terms will be present in the expansion.

The first two terms in this expansion are given by
\begin{equation}\label{g0}
  G_N^{(0)}=\lambda^N\sum_{n=1}^{\infty}\int
  \frac{d^{\D}p}{(2\pi)^{\D}}\left(\frac{1}{p^2+M^2n^2}\right)^N,
\end{equation}
for the zeroth term and by
\begin{eqnarray}\label{g2}
  &&\qquad G_N^{(2)}= \\ \nonumber
  &&\frac{\lambda^N}{2N!}\sum_{n=1}^{\infty}\int
  \frac{d^{\D}p}{(2\pi)^{\D}}\left(\sum_i Q_i^\mu Q_i^\nu
  \left(\frac{1}{p^2+M^2n^2}\right)^{N-1}
  \frac{\pd^2}{\pd p_\mu \pd p_\nu}\left(\frac{1}{p^2+M^2n^2}\right)\right. +
  \\ \nonumber
  &&\left. \sum_{i\neq j}Q_i^\mu Q_j^\nu \left(\frac{1}{p^2+M^2n^2}\right)^{N-2}
  \frac{\pd }{\pd p_\mu} \left(\frac{1}{p^2+M^2n^2}\right)
  \frac{\pd }{\pd p_\nu}\left(\frac{1}{p^2+M^2n^2}\right)\right)+
  \\ \nonumber
  && \qquad \qquad + \text{permutations}(q_1,\dots , q_N),
\end{eqnarray}
for next one.

Computation of the integrals (\ref{g0}, \ref{g2}) yields,
respectively,
\begin{eqnarray}\label{f0}
  G_N^{(0)}\equiv \frac{\lambda^N}{(2\sqrt{\pi})^{\D}(N-1)!}
  M^{\D-2N}f_N^{(0)}=\\ \nonumber
  \frac{\lambda^N}{(2\sqrt{\pi})^{\D}(N-1)!} M^{\D-2N}
  \zeta(2N-\D)\Gamma(N-\D/2),
\end{eqnarray}
and
\begin{eqnarray}\label{f2}
    G_N^{(2)}\equiv \frac{\lambda^N}{(2\sqrt{\pi})^{\D}(N-1)!}
  M^{\D-2(N+1)}f_N^{(2)}=\\ \nonumber
  \frac{\lambda^N}{(2\sqrt{\pi})^{\D}(N-1)!}
  M^{\D-2(N+1)}\frac1{2N}\zeta(2N+2-\D)\Gamma(N+2-\D/2)P_{(2)}(q),
\end{eqnarray}
where
\begin{equation}\label{p2}
   P_{(2)}(q)=\frac{1}{N!}\sum_{\text{perm}(q_1\dots
   q_N)}\left(4\sum_{i,j} (Q_i\cdot
   Q_j)\left(\frac{1}{N+1}\right)-
   \sum_{i}Q_i^2\left(\frac12+\frac{3}{N+1}\right)\right).
\end{equation}

Next terms can be computed in a similar manner, but we are
interested only in those terms which are of the type contained in
the classical Lagrangian (\ref{d+1}).

%-------------------------------------------------------------------------
\section{Zeta and Gamma Function Properties}

In this Appendix we review some properties of Gamma and Zeta
functions borrowed from \cite{guide}, which are relevant for us.

\subsection*{Gamma function}

For $\RE s>0$ Gamma function is defined by the integral
\begin{equation}\label{gamma}
  \Gamma (s) =\int_0^\infty t^{s-1}e^{-t}dt.
\end{equation}

It can be analytically continued to the values $\RE\leq0$. On the
negative part of real axis, however this function has
singularities (simple poles) at negative integer point as well as
at the zero point.

We are mainly interested in function's behaviour at zero point:
\begin{equation}\label{gamma(0)}
  \Gamma(\varepsilon)=\frac1{\varepsilon}-\gamma + O(\varepsilon),
\end{equation}
where $\gamma\approx 0.577216$ is the Euler constant. At negative
points poles will be cancelled by zeroes of
\subsection*{Zeta function} Zeta function $\zeta (s)$ is defined
by the series
\begin{equation}\label{zeta}
  \zeta (s)=\sum_{n=1}^{\infty} n^{-s}.
\end{equation}
This defines zeta-function for $\RE s>1$. Analytic continuation to
$\RE s\leq 1$ has a pole at $s=1$. Near the singularity point
$\zeta(s)$ behaves like
\begin{equation}\label{sing}
  \zeta
  (s)=\frac{1}{s-1}+\sum_{n=0}^{\infty}\frac{(-1)^n}{n!}\gamma_n(s-1)^n,
\end{equation}
where
\begin{equation}\label{g}
  \gamma_n
  =\lim_{m\rightarrow\infty}\left\{\sum_{k=1}^{m}\frac{(\log
  k)^n}{k}-\frac{(\log m)^{n+1}}{n+1}\right\}.
\end{equation}

Zeta-function, also, have zeroes at even negative integer values
of $s$:
\begin{equation}\label{zzero}
  \zeta (-2n)=0, \qquad n \in \z.
\end{equation}
These zeroes just compensate all poles of Gamma function in
product
\begin{equation}\label{factor}
  \zeta (2s)\Gamma (s),
\end{equation}
except one for $s=0$ since
\begin{equation}\label{zero}
  \zeta (0)= -1/2.
\end{equation}

Thus two singular points are $s=0$ and $s=\frac12$ and
\begin{equation}\label{z}
   \zeta (2\varepsilon)\Gamma
   (\varepsilon)=-\frac1{2\varepsilon}+\frac12\gamma- \log
   2\pi
   +O(\varepsilon),
\end{equation}
and
\begin{equation}\label{h}
     \zeta (1+\varepsilon)\Gamma
   (1/2+\varepsilon/2)=\frac{\sqrt{\pi}}{\varepsilon}+
  \sqrt{\pi}\left(\gamma+ \frac12\psi(1/2)\right)+O(\varepsilon),
\end{equation}
where $\psi (1/2)\approx -1.96351$.

Also,
\begin{equation}\label{beta}
  \int_0^\infty\left(\frac{1}{t+M^2n^2}\right)^\beta t^\alpha
  dt=(M^2n^2)^{\alpha+1-\beta}
  \frac{\Gamma(\alpha+1)\Gamma(\beta-\alpha-1)}{\Gamma(\beta)},
\end{equation}
is used in Appendix A.

% ------------------------------------------------------------------------

%\bibliographystyle{unsrt}
%\bibliography{qKKc}

\end{document}